\documentclass[journal=tches,preprint]{iacrtrans}

\usepackage{hyperref}
\usepackage{xspace}
\usepackage{import}
\usepackage{listings}
\usepackage{booktabs}
\usepackage{amssymb}
\usepackage{todonotes}
\usepackage{amsmath}
\usepackage{upgreek}
\usepackage{comment}
\usepackage{breakurl}

\usepackage[normalem]{ulem}

\usepackage[framemethod=tikz]{mdframed}
\usepackage{caption}
\usepackage[font={footnotesize,md}]{caption} 
\usepackage[nolist]{acronym} 

\usepackage{cleveref}
\Crefname{lstlisting}{listing}{listings}
\Crefname{lstlisting}{Listing}{Listings}

\usepackage[]{todonotes}
\usepackage{enumitem}

\let\oldtodo\todo
\renewcommand{\todo}[1]{\oldtodo[inline]{#1}}

\ifdefined\withchanges
\newcommand{\changed}[1]{{\color{blue}#1}}
\newcommand{\removed}[1]{{\color{red}\sout{#1}}}
\else
\newcommand{\changed}[1]{{#1}}
\newcommand{\removed}[1]{{}}
\fi

\author{Zitai Chen \and David Oswald}
\institute{
  University of Birmingham, Birmingham, UK, \email{Z.Chen@pgr.bham.ac.uk}
  \and
  University of Birmingham, Birmingham, UK, \email{d.f.oswald@bham.ac.uk}
}

\title{PMFault: Faulting and Bricking Server CPUs through Management Interfaces}
\subtitle{Or: A Modern Example of Halt and Catch Fire}

\newcommand{\code}[1]{{\texttt{#1}}}

\newcommand{\etal}{et~al.\xspace} 
\newcommand{\ie}{\textit{i.e.},\ } 
\newcommand{\eg}{e.g.,\ } 
\newcommand{\cf}{cf.\ } 
\newcommand{\pvCVE}{CVE-2019-11157\xspace}

\newcommand{\tSuper}{E3-1220V6-X11SSL-CF\xspace}
\newcommand{\tSuperMobo}{Supermicro X11SSL-CF\xspace}
\newcommand{\tAsRockServer}{ASRock E3C246D4I-2T\xspace}
\newcommand{\tSuperMoboNew}{Supermicro X12DPi-NT6\xspace}

\newcommand{\pmattackname}{PMFault\xspace}
\newcommand{\pmdetect}{PMBusDetect\xspace}

\begin{document}
\begin{acronym}[ASCII]
 \setlength{\itemsep}{0.2em}
 \acro{PCI-E}{Peripheral Component Interconnect Express}
 \acro{BMC}{Baseboard Management Controller}
 \acro{I2C}{Inter-Integrated Circuit}
 \acro{VRM}{Voltage Regulator Module}
 \acro{PMBus}{Power Management Bus}
 \acro{SMBus}{System Management Bus}
 \acro{SVID}{Serial Voltage Identification}
 \acro{KCS}{Keyboard Controller Style}
 \acro{IPMI}{Intelligent Platform Management Interface}

\acroplural{CPA}[CPAs]{Correlation Power Analyzes}
\acroplural{DUT}[DUTs]{Devices Under Test}
\acroplural{OS}[OSes]{Operating Systems}
\acroplural{RTOS}[RTOSes]{Real-Time Operating Systems}
\acro{3DES}{Triple DES}
\acro{AES}{Advanced Encryption Standard}
\acro{ANF}{Algebraic Normal Form}
\acro{APB}{Advanced Peripheral Bus}
\acro{API}{Application Programming Interface}
\acro{ARX}{Addition, Rotate, XOR}
\acro{ASIC}{Application Specific Integrated Circuit}
\acro{ASK}{Amplitude-Shift Keying}
\acro{AXI}{Advanced Extensible Interface}
\acro{BCM}{Body Control Module}
\acro{BL}{Bootloader Enable}
\acro{BOR}{Brown-Out Reset}
\acro{BPSK}{Binary Phase Shift Keying}
\acro{CAN}{Controller Area Network}
\acro{CBC}{Cipher Block Chaining}
\acro{CBS}{Critical Bootloader Section}
\acro{CFG}{Control Flow Graph}
\acro{CGM}{Continuous Glucose Monitoring System}
\acro{CMOS}{Complementary Metal Oxide Semiconductor}
\acro{CMT}{Compare and Match Timer}
\acro{COPACOBANA}{Cost-Optimized Parallel Code Breaker and Analyzer}
\acro{CPA}{Correlation Power Analysis}
\acro{CPU}{Central Processing Unit}
\acro{CRC}{Cyclic Redundancy Check}
\acro{CRP}{Code Readout Protection}
\acro{CSTF}{CoreSight Trace Funnel}
\acro{CTR}{Counter \acroextra{(mode of operation)}}
\acro{DC}{Direct Current}
\acro{DDS}{Digital Direct Synthesis}
\acro{DES}{Data Encryption Standard}
\acro{DFA}{Differential Fault Analysis}
\acro{DFT}{Discrete Fourier Transform}
\acro{DMA}{Direct Memory Access}
\acro{DoS}{Denial-of-Service}
\acro{DPA}{Differential Power Analysis}
\acro{DRAM}{Dynamic Random-Access Memory}
\acro{DRM}{Digital Rights Management}
\acro{DSO}{Digital Storage Oscilloscope}
\acro{DSP}{Digital Signal Processing}
\acro{DST}{Digital Signature Transponder}
\acro{DUT}{Device Under Test}
\acro{ECB}{Electronic Code Book}
\acro{ECC}{Elliptic Curve Cryptography}
\acro{ECU}{Electronic Control Unit}
\acro{EDE}{Encrypt-Decrypt-Encrypt \acroextra{(mode of operation)}}
\acro{EEPROM}{Electrically Erasable Programmable Read-Only Memory}
\acro{ELF}{Executable and Linkable Format}
\acro{EM}{Electro-Magnetic}
\acro{ETF}{Embedded Trace Funnel}
\acro{ETR}{Embedded Trace Router}
\acro{FCC}{Federal Communications Commission}
\acro{FD}{Feedback Disassembly}
\acro{FFT}{Fast Fourier Transform}
\acro{FIR}{Finite Impulse Response}
\acro{FIVR}{Fully Integrated Voltage Regulator}
\acro{FPGA}{Field Programmable Gate Array}
\acro{FSK}{Frequency Shift Keying}
\acro{GCM}{Gateway Control Module}
\acro{GIAnT}{Generic Implementation ANalysis Toolkit}
\acro{GMSK}{Gaussian Minimum Shift Keying}
\acro{GPIO}{General Purpose I/O}
\acro{GPR}{General Purpose Register}
\acro{GPT}{General Purpose Timer}
\acro{HAL}{Hardware Abstraction Layer}
\acro{HDL}{Hardware Description Language}
\acro{HD}{Hamming Distance}
\acro{HF}{High Frequency}
\acro{HMAC}{Hash-based Message Authentication Code}
\acro{HW}{Hamming Weight}
\acro{I2C}{Inter-Integrated Circuit}
\acro{IC}{Instrument Cluster}
\acro{IC}{Integrated Circuit}
\acro{ID}{Identifier}
\acro{IIR}{Infinite Impulse Response}
\acro{IL}{Intermediate Language}
\acro{IoT}{Internet of Things}
\acro{IP}{Intellectual Property}
\acro{IRQ}{Interrupt Request}
\acro{IR}{Intermediate Representation}
\acro{ISA}{Instruction Set Architecture}
\acro{ISM}{Industrial, Scientific, and Medical \acroextra{(frequencies)}}
\acro{ISR}{Interrupt Service Routine}
\acro{IV}{Initialization Vector}
\acro{JTAG}{Joint Test Action Group}
\acro{LCP}{Longest Common Prefix}
\acro{LFSR}{Linear Feedback Shift Register}
\acro{LF}{Low Frequency}
\acro{LQI}{Link Quality Indicator}
\acro{LSByte}{Least Significant Byte}
\acro{LSB}{Least Significant Bit}
\acro{LTE}{Long Term Evolution}
\acro{LUT}{Look-Up Table}
\acro{MAC}{Message Authentication Code}
\acro{MCU}{Microcontroller Unit}
\acro{MF}{Medium Frequency}
\acro{MITM}{Man-In-The-Middle}
\acro{MMIO}{Memory-mapped I/O}
\acro{MMU}{Memory Management Unit}
\acro{MSByte}{Most Significant Byte}
\acro{MSB}{Most Significant Bit}
\acro{MSK}{Minimum Shift Keying}
\acro{MSR}{Model Specific Register}
\acro{muC}[$\mathrm{\upmu C}$]{Microcontroller}
\acro{muC}{Microcontroller}
\acro{NAS}{Non-Access Stratum}
\acro{NFC}{Near Field Communication}
\acro{NLFSR}{Non-Linear Feedback Shift Register}
\acro{NLF}{Non-Linear Function}
\acro{NRZ}{Non-Return-to-Zero \acroextra{(encoding)}}
\acro{NVM}{Non-Volatile Memory}
\acro{OCD}{On-Chip Debug}
\acro{OOK}{On-Off-Keying}
\acro{OP}{Operational Amplifier}
\acro{OS}{Operating System}
\acro{OTP}{One-Time Password}
\acro{PCB}{Printed Circuit Board}
\acro{PC}{Personal Computer}
\acro{PhD}{Patiently hoping for a Degree}
\acro{PKES}{Passive Keyless Entry and Start}
\acro{PKE}{Passive Keyless Entry}
\acro{PKI}{Public Key Infrastructure}
\acro{PMBus}{Power Management Bus}
\acro{PoC}{Proof-of-Concept}
\acro{POR}{Power-On Reset}
\acro{PPC}{Pulse Pause Coding}
\acro{PRNG}{Pseudo-Random Number Generator}
\acro{PSK}{Phase Shift Keying}
\acro{PTM}{Program Trace Module}
\acro{PWM}{Pulse Width Modulation}
\acro{RAPL}{Running Average Power Limit}
\acro{RDP}{Read-out Protection}
\acro{RE}{Reverse Engineering}
\acro{RFID}{Radio Frequency IDentification}
\acro{RF}{Radio Frequency}
\acro{RKE}{Remote Keyless Entry}
\acro{RNG}{Random Number Generator}
\acro{ROM}{Read Only Memory}
\acro{ROP}{Return-Oriented Programming}
\acro{RRC}{Radio Resource Control}
\acro{RSA}{Rivest Shamir and Adleman}
\acro{RSSI}{Received Signal Strength Indicator}
\acro{RTL}{Register Transfer Language}
\acro{RTOS}{Real-Time Operating System}
\acro{SCA}{Side-Channel Analysis}
\acro{SDR}{Software Defined Radio}
\acro{SDR}{Software-Defined Radio}
\acro{SGX}{Software Guard Extensions}
\acro{SHA-1}{Secure Hash Algorithm 1}
\acro{SHA-256}{Secure Hash Algorithm 2 (256-bit version)}
\acro{SHA}{Secure Hash Algorithm}
\acro{SHF}{Superhigh Frequency}
\acro{SMA}{SubMiniature version A \acroextra{(connector)}}
\acro{SMBus}{System Management Bus}
\acro{SMT}{Satisfiability Modulo Theories}
\acro{SNR}{Signal to Noise Ratio}
\acro{SoC}{System on Chip}
\acro{SoC}{System-on-Chip}
\acro{SEV}{Secure Encrypted Virtualization}
\acro{SP}{Secure Processor \acroextra{(AMD)}}
\acro{SPA}{Simple Power Analysis}
\acro{SPI}{Serial Peripheral Interface}
\acro{SPOF}{Single Point of Failure}
\acro{SUT}{System Under Test}
\acro{SVID}{Serial Voltage Identification}
\acro{SWD}{Serial Wire Debug}
\acro{TCB}{Trusted Computing Base}
\acro{TCU}{Telematics Control Unit}
\acro{TEE}{Trusted Execution Environment}
\acro{TMDTO}{Time-Memory-Data Tradeoff}
\acro{TMTO}{Time-Memory Tradeoff}
\acro{TPIU}{Trace Port Interface Unit}
\acro{TZ}[TrustZone]{TrustZone}
\acro{UART}{Universal Asynchronous Receiver Transmitter}
\acro{UCODE}{Microcode}
\acro{UDS}{Unified Diagnostic Services}
\acro{UHF}{Ultra High Frequency}
\acro{UID}{Unique Identifier}
\acro{USB}{Universal Serial Bus}
\acro{USRP2}{Universal Software Radio Peripheral (version 2)}
\acro{USRP}{Universal Software Radio Peripheral}
\acro{VHDL}{VHSIC (Very High Speed Integrated Circuit) Hardware Description Language}
\acro{VHF}{Very High Frequency}
\acro{VLE}{variable length encoded}
\acro{VLF}{Very Low Frequency}
\acro{VR}{Voltage Regulator}
\acro{VXE}{Virtual Execution Environment}
\acro{WLAN}{Wireless Local Area Network}
\acro{XOR}{Exclusive OR}
\acroplural{CFG}[CFGs]{Control-Flow Graphs}
\acro{VID}{Voltage Identifier}
\acro{OCP}{Over Current Protection}
\acro{Intel ME}{Intel Management Engine}
\end{acronym}

\maketitle

\begin{abstract}
Apart from the actual CPU, modern server motherboards contain other auxiliary components, for example voltage regulators for power management. Those are connected to the CPU and the separate Baseboard Management Controller (BMC) via the I2C-based PMBus. 
In this paper, using the case study of the widely used Supermicro X11SSL motherboard, we show how remotely exploitable software weaknesses in the BMC (or other processors with PMBus access) can be used to \removed{mount} \changed{access the PMBus and then perform} hardware-based fault injection attacks on the main CPU. \changed{The underlying weaknesses include insecure firmware encryption and signing mechanisms, a lack of authentication for the firmware upgrade process and the IPMI KCS control interface, as well as the motherboard design (with the PMBus connected to the BMC and SMBus by default).} 
First, we show that undervolting through the PMBus allows breaking the integrity guarantees of SGX enclaves, bypassing Intel's countermeasures against previous undervolting attacks like Plundervolt/V0ltPwn. Second, we experimentally show that overvolting outside the specified range has the potential of permanently damaging Intel Xeon CPUs, rendering the server inoperable. We assess the impact of our findings on other server motherboards made by Supermicro and ASRock.
\changed{Our attacks, dubbed \pmattackname,} can be carried out by a privileged software adversary and do not require physical access to the server motherboard or knowledge of the BMC login credentials.
We responsibly disclosed the issues reported in this paper to Supermicro and discuss possible countermeasures at different levels. \changed{To the best of our knowledge, the 12th generation of Supermicro motherboards, which was designed before we reported \pmattackname to Supermicro, is not vulnerable. }

\removed{\noindent\textbf{Embargo notice:} Supermicro have requested for this paper to be treated as \emph{embargoed} until at least November 2022. }
\end{abstract}

\section{Introduction}
In recent years, the security implications of software-exposed power and clock management features have received substantial attention by the research community. Several attacks including CLKSCREW~\cite{Tang17}, Plundervolt~\cite{murdock2019plundervolt}, V0ltPwn~\cite{Kenjar20}, and VoltJockey~\cite{Qiu19_sgx} showed that undervolting or overclocking from software can be used to inject faults (\eg bitflips) into computations and break \acp{TEE} like Intel \changed{\ac{SGX}} and ARM TrustZone. Subsequent attacks like VoltPillager~\cite{Chen21} and the work by Buhren \etal\ \cite{Buhren21} showed that similar attacks can be mounted with direct access to the computer hardware, physically connecting to the control interface of the \ac{VR}.

In particular, Chen \etal targeted the \ac{SVID} interface used by Intel CPUs to set the desired supply voltage. However, apart from \ac{SVID}, many systems, in particular servers, support a second interface, the so-called \ac{PMBus}, to control the \ac{VRM}. 
\ac{PMBus} is an open standard for digital power management \cite{pmbus13} and has been adopted by more than 40 companies. It is based on the \ac{I2C} bus and offers monitoring features apart from voltage and current control.

Another component usually presents on server motherboards is the \ac{BMC}. This chip, intended to remotely manage the server even if \eg the main CPU has crashed or is powered down, has connections to several buses and chips on the motherboard, including the \ac{I2C} bus on which the \ac{VRM} resides. 

\changed{Previous research on x86 platforms has focused on the software-hardware interface provided by the \ac{CPU} itself and on the security within the perimeter of each individual component, \eg the \ac{BMC}~\cite{Perigaud18} or \ac{Intel ME}~\cite{me-ring3-root, me-2017-silentbob, me-5705}. There is a lack of board-level security analysis that reviews the system and motherboard design and interactions between the different components: even if an individual part of the system is secure within its individual threat model, the combination of it with other parts can cause security risks.}
\changed{In our \pmattackname attacks,} the privileged position of \changed{the} \ac{BMC}, combined with \changed{its} large attack surface, makes \removed{BMC} \changed{it} interesting from an adversary's perspective to exploit vulnerabilities \changed{of the system} via power management features.

\subsection{Our Contribution}

 \changed{Our main contributions in this paper are:}

\emph{\changed{\ac{PMBus}-based under/overvolting against server platforms:}}
\removed{In this paper,} We first analyse the \ac{VRM}\removed{-related server} management interface at the hardware level. We discovered that the semi-standardised \ac{PMBus} can be used to control the CPU voltage. Using the case study of a widely-used server motherboard, the \tSuperMobo, we explore this attack surface and show that software vulnerabilities in the \ac{BMC} (or another programmable chip connected to the \ac{PMBus}) can have severe consequences for the security and safety of the server platform. \changed{To determine if the vulnerabilities can affect other server motherboards, we also investigated the \ac{PMBus} connections and usage on an \tAsRockServer and a \tSuperMoboNew.}

\emph{\changed{\ac{PMBus} access through \ac{BMC} exploits:}}
We then study the \ac{BMC} firmware and---based on prior work in \cite{eclypsium_2018,ipmi-fw-tool,smcbmc-tool}---found that it can indeed be exploited to send arbitrary \ac{PMBus} commands to control the voltage of the CPU.  More precisely, several software vulnerabilities in the \ac{BMC}, including incorrect firmware encryption and signing mechanisms, a lack of authentication for firmware upgrades and control interfaces, an attacker can manipulate the \ac{CPU} voltage remotely because the \ac{PMBus} is connected to the \ac{BMC} and the \ac{SMBus} by default. 

\emph{\changed{\ac{PMBus}-based undervolting against \ac{SGX} enclaves:}} With this, we observed the same faults as with \changed{Plundervolt/V0ltPwn (\pvCVE)}, including for code running inside an \changed{\acs{SGX} enclave}. As the \ac{BMC} has an independent, \changed{external} flash \changed{chip} for its firmware, \ac{SGX} attestation currently \emph{does not} have the ability to verify its status. Crucially, because the software \changed{voltage-control} interface in \ac{MSR} \texttt{0x150} is not used, Intel's fix for \changed{\pvCVE} does not address this attack. 

\emph{\changed{Permanent denial-of-service through overvolting:}}
\removed{Second,} We \changed{also} discovered a \changed{novel} overvolting attack: by sending a certain sequence of \ac{PMBus} commands, we can set the CPU voltage outside the specification (as high as 2.84\,V) and permanently brick the Xeon CPU used in our experiments. 
\removed{Third, to find out if the vulnerabilities found in this paper can affect other server motherboards, we also investigated the \ac{PMBus} connection on \tAsRockServer and \tSuperMoboNew.}

\emph{\changed{Countermeasures and mitigations:}}
\changed{Finally, we develop the \pmdetect tool for detecting if the \ac{VRM} is connected to the \ac{PMBus}, and then discuss countermeasures and challenges in securing server platforms. Importantly, we point out that \acp{TEE} like \ac{SGX} must not only rely on the security of the CPU itself, but also of that of management components the hardware design of the platform.} \removed{In summary, our main contributions are:}

\ifdefined\withchanges

\begin{enumerate}[leftmargin=15pt, topsep=1pt, partopsep=4pt, itemsep=-3pt, parsep=3pt]

	\item \removed{We demonstrate the first \ac{PMBus}-based under/overvolting attacks against server platforms without requiring additional hardware.}
	
	\item \removed{We show that remote and local privileged attackers can use vulnerabilities in \acp{BMC} to gain full control of the CPU voltage.}
	
	\item \removed{We show that our undervolting attack compromises the integrity of \ac{SGX} computations and that  Intel's previous countermeasures against software-based undervolting (\pvCVE) does not prevent the attack.}
	
	\item \removed{We demonstrate a novel overvolting attack, which can destroy the server CPU from software and thus lead to permanent denial-of-service.}
	
	\item \removed{Finally, we develop the \pmdetect tool for detecting if the \ac{VRM} is connected to \ac{PMBus}, and then discuss countermeasures and challenges in securing server platforms.}  
	
\end{enumerate}
\else
\fi

\removed{We will release details of our experiments and source after the publication of the paper. For the reviewers only, we provide a copy under the following URL:} 

\removed{\url{https://mega.nz/folder/wo5QRIzA\#sZuU1x9GTD2OJBxWoSwZjA }}

\changed{The details of our experiments and source code can be found at: \url{https://github.com/zt-chen/PMFault}. CVE number CVE-2022-43309 has been reserved for PMFault.}

\subsection{Adversary Model}
\label{sec:adversary}
In this paper, we assume a privileged software attacker, \ie who has obtained \code{root} on the host CPU. This is the standard adversary model in the case of \acp{TEE} like \ac{SGX}, and is also realistic in the case of overvolting to permanently destroy the CPU, which could be \eg exploited by ransomware with \code{root} rights.
Notably, our attacks do not require physical access (for additional hardware to be added to the system) and can thus be conducted remotely \eg over SSH.

\subsection{Responsible Disclosure}
We have responsibly disclosed our findings to Intel and Supermicro in April 2022. We discussed the details of our methods in several calls with Supermicro, and they acknowledge the existence of the issue and are looking into deploying fixes for their 11th generation products like the \tSuperMobo. Supermicro highlighted that the attacks do not replicate on their 12th generation, which \eg include secure boot and update for the \ac{BMC} and filtering on \ac{PMBus} commands. \changed{Both of these features break the attack chains described in the paper.} Intel considered the issue in the context of their own server motherboards and did not find them vulnerable. Intel did not comment on the impact on \ac{SGX}.

\removed{\noindent\textbf{Embargo notice:} Supermicro have requested for this paper to be treated as \emph{embargoed} until at least November 2022. Please treat this paper as confidential.}

\subsection{Related Work}
Since Boneh \etal's seminal work on fault injection~\cite{BonehDemilloLipton97}, the research community has devoted substantial efforts to investigating fault attacks and developing according countermeasures (\cf \eg\ \cite{Bar2006sorcerer} for an overview). 

\paragraph{Software-based Fault Injection}
Often, fault injection was considered a technique limited to attacks with physical access to the target. However, with the discovery of the Rowhammer effect~\cite{Kim2014}, it was shown that faults can also be injected from software (through specific memory access patterns in the case of Rowhammer). Then, in 2017, Tang \etal showed that the clock management features of ARM processors can be exploited to inject faults into computations shielded inside a \ac{TEE} like ARM TrustZone~\cite{Tang17}. 
Similarly, Plundervolt, V0ltPwn, and VoltJockey~\cite{murdock2019plundervolt,Kenjar20,Qiu19_sgx} (all tracked via \pvCVE) use the software-exposed voltage control \ac{MSR} in Intel processors to break the integrity guarantees of \ac{SGX} enclaves. In response, Intel deployed a microcode update that disables the undervolting interface in \ac{MSR} \texttt{0x150} and allows remote parties to verify correct configuration through \ac{SGX}'s remote attestation. Thus, purely software-based undervolting attacks against Intel processors were considered no longer possible. 

\paragraph{Hardware-based Fault Injection on \acp{TEE}}
The second generation of undervolting attacks on \acp{TEE} like \ac{SGX} and AMD \ac{SEV} require physical access to the target motherboard. In the case of VoltPillager~\cite{Chen21}, the adversary attaches two wires to the data and clock lines of the \ac{SVID} bus and can then control the \ac{VRM} external to the CPU, enabling undervolting even if Intel's microcode fixes for \pvCVE are installed. For AMD \ac{SEV}, the adversary does not glitch the actual CPU, but the separate security co-processor, the AMD \ac{SP}~\cite{Buhren21}. The adversary then proceeds to upload custom firmware to the \ac{SP} to leak memory encryption keys and also endorsement secrets, which ultimately enable attacks without permanent physical access.

\paragraph{Security of servers and \acp{BMC}}
Independent of hardware-based attacks, the security of server platforms has received attention in the research community and wider society. In 2018, Bloomberg published a---since widely disproven---article that \emph{incorrectly} claimed the inclusion of small backdoor chips on Supermicro motherboards~\cite{Bloomberg18}. However, at the same time, researchers at Eclypsium showed that it is indeed possible to maliciously manipulate the \ac{BMC} firmware of Supermicro motherboards from 8th to 11th generation~\cite{eclypsium_2018}, without the need to add a hardware implant. They also demonstrated how flashing corrupted \ac{BMC} firmware can ``brick'' the server system by preventing it to boot. 

Niewöhner~\cite{smcbmc-tool} subsequently published a tool to exploit the (weak) firmware encryption of Supermicro \acp{BMC}.
Other work, for example by Waisman \etal~\cite{Waisman18} and Périgaud \etal~\cite{Perigaud18}, has shown that software weaknesses in \acp{BMC} are not limited to Supermicro motherboards, but also applied to Dell, HP, and Lenovo systems. 

However, the implications of direct access to the \ac{PMBus} from a compromised \ac{BMC} have not been deeply studied to our knowledge.

\subsection{Paper Outline}
The remainder of this paper is structured as follows: in \Cref{sec:pmbus_intro}, we review the \ac{PMBus} protocol and analyse its specific implementation and usage on Supermicro motherboards. Then, in \Cref{sec:BMC}, we describe Supermicro's \ac{BMC} implementation and methods to modify the firmware. In \Cref{sec:practical}, we experimentally investigate how a compromised \ac{BMC} can interact with the \ac{VRM} through the \ac{PMBus}. We then use this to develop over/undervolting attacks in \Cref{sec:attacks}, before concluding in \Cref{sec:conclusion}.

\section{Analysis of \acl{PMBus}}\label{sec:pmbus_intro}
\changed{We started our work by analysing how the \ac{PMBus} is used on practical server motherboards.} \ac{PMBus} is an interface that is used to control the \ac{VRM}, supplying the power to the CPU. The most recent public available specification is version 1.3~\cite{pmbus13}. This specification standardises the physical interface, packet structure, and command set of the \ac{PMBus}. However, some commands are left as ``manufacturer specified'', so that each \ac{VRM} manufacturer can have a slightly different implementation of the command set. This matches what we found during our investigation of the MP2955 \ac{VRM} on the \tSuperMobo platform described in the following. 

\subsection{Experimental Setup}\label{sec:expsetup}

\begin{figure}[h]
  \centering
  \resizebox{0.65\columnwidth}{!}{

    \def\svgwidth{\textwidth}
    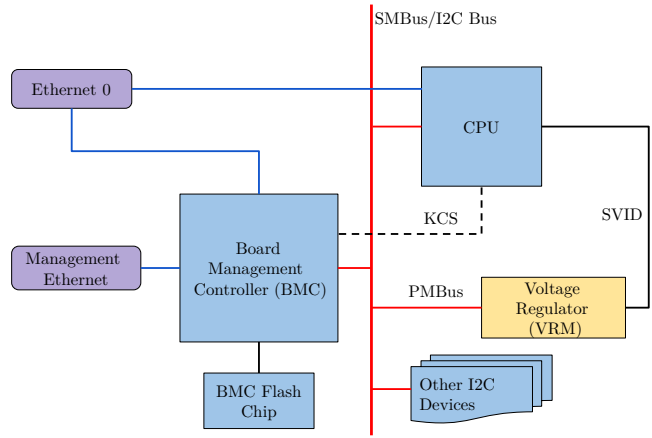

  }
  \caption{Overview of the connections on the server motherboard.\label{fig:sysconnection}}
\end{figure}

We carried out initial experiments with an Intel Xeon E3-1220 v6 (CPU family: 6, model: 158, microcode version: 0xea) on a \tSuperMobo Rev 1.01 motherboard (\ac{BMC} microcontroller ASPEED AST2400, firmware revision 01.63, BIOS version: 2.4).We used 64-bit Ubuntu 18.04.3 LTS with a stock 5.4.0-107-generic kernel, Intel SGX driver V2.11.0, and Intel SGX-SDK V2.15.100.3. We refer to this system as \tSuper throughout the paper. \changed{An overview of the server motherboard representative for Supermicro's 11th generation products is shown in \Cref{fig:sysconnection}. The target of the \pmattackname attack is an Intel CPU with \ac{SGX} technology.} As mentioned, our actual attacks do not require additional hardware or physical access to the system, though we soldered some wires to the motherboard during the analysis phase.

\removed{An overview of the server motherboard, which is representative for Supermicro's 11th generation products, is shown in Figure 1. The target of our attack, \pmattackname, is an Intel CPU with \ac{SGX} technology.} On Intel platforms, the voltage of the CPU is controlled by an external  \ac{VRM} \ac{IC}. The CPU connects to the \ac{VRM} via the \ac{SVID} bus to control the voltage supplied by it. This interface for CPU voltage control is present on all desktop and server motherboard\changed{s}.

However, server \acp{VRM}---including the \tSuperMobo---often have an additional \ac{I2C}-based communication interface called \ac{PMBus}. This interface allows \eg overclocking or fine-tuning of the CPU voltage. One of the crucial steps in the \pmattackname is to get access to this interface and understand the communication protocol, so that we gain full control of the CPU voltage. 

One of the design issues we found on our server motherboard is that the \ac{PMBus} can be directly connected to the more general \ac{SMBus}. There are various components on the system on that bus, including the \ac{CPU}, \ac{BMC}, and other \ac{I2C} devices. A compromise of any of these components leads to the takeover of \ac{PMBus} and thus control of the CPU voltage. 

In this paper, we use the \ac{BMC} as the starting point of the attack, as it commonly exists on server platforms. In order to analyse the attack surface of the \ac{BMC}, we further investigated its connection and hardware design on the \tSuperMobo. First, we found that its firmware is stored in a \ac{SPI} flash chip, separate from the BIOS flash. 
We also found there are two Ethernet ports on the system for communication with the \ac{BMC}: one is called ``Management Ethernet'' and is dedicated for server management. The other port can be shared between CPU and \ac{BMC} so that devices on this Ethernet port can communicate with both CPU and \ac{BMC}. Finally, the \ac{BMC} also has a \ac{KCS} interface that enables direct access from the \ac{OS} running on the CPU. These management interfaces open a large attack surface on the \ac{BMC}, and make remote attacks possible. 

\subsection{Protocol Structure}

\begin{figure}[h]
  \centering
  \resizebox{0.8\columnwidth}{!}{
    \def\svgwidth{1.1\textwidth}
    \footnotesize
    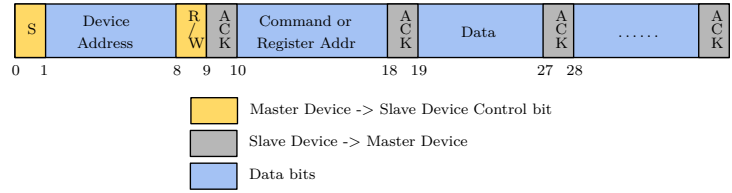

  }
  \caption{PMBus protocol structure\label{fig:pmbusProtocol}}
\end{figure}

\changed{To be able to eavesdrop and forge \ac{PMBus} commands, knowledge of the protocol structure shown in \Cref{fig:pmbusProtocol} is necessary.}  \changed{The \ac{PMBus}} is an \ac{I2C}-based protocol (\changed{with clock speed of 100\,kHz--1\,MHz} and an open-drain data pin) and uses a master-slave communication mechanism. The master device can query or change the setting of the slave device. Each slave device is assigned a unique 7-bit device address. 

The master device first sends a starting bit to initiate a transmission. During transmission, every group of 9 bits forms a segment, with the 9th bit indicating ACK (0) or NACK (1) for every 8 bits received. The starting bit and the (N)ACK mechanism are handled at hardware level and do not need to be handled manually. 

The first segment is always sent by the master. The first 7 bits are the address of the target slave, and the 8th bit indicates whether this transmission is a read (1) or write (0). The second segment is the register address to operate on. In the \ac{PMBus} specification, this segment is called the \emph{\ac{PMBus} command}. The segments after the second one contain the data read from or written to the register.

\paragraph{Interaction between \ac{PMBus} and \ac{SVID}}
Although the functionality of the \ac{PMBus} protocol is similar to \ac{SVID}, they have different specifications for the digital signal interface and command sets. A \ac{VRM} can have both \ac{SVID} and \ac{PMBus} interfaces, with the \ac{SVID} interface directly connected to the CPU and the \ac{PMBus} interface connected to the \ac{SMBus}. Both interfaces can be used to control the voltage of the CPU, and some implementations of the \ac{PMBus} specification also have commands to override the voltages set through the \ac{SVID} interface.

\subsection{\acs{PMBus} Commands}

\changed{For an adversary to communicate with the \ac{VRM} and \eg configure voltage levels, they also need to know the specific \ac{PMBus} commands.} As mentioned, the \ac{PMBus} specification allows manufacturers to have custom implementations of \ac{PMBus} commands. The \tSuper motherboard features an Monolithic Power MP2955 voltage regulator. To understand the \ac{PMBus} implementation of this \ac{VRM}, we first started looking for its datasheet, but unfortunately, found that it is not publicly available. However, on the Monolithic Power website\footnote{\url{https://www.monolithicpower.com/}}, we found the datasheet of an alternative \ac{VRM} (MP2965)~\cite{mp2965}. As both chips are manufactured by the same company, we used this datasheet as a reference and starting point to discover the available \ac{PMBus} commands by analysing\removed{sniffing} the \ac{PMBus} traffic on the \tSuperMobo. 

\changed{We found the relevant \ac{PMBus} commands by} reading and analysing the response \changed{(ACK or NACK)} of the registers, and validating \changed{found commands} according to the \ac{PMBus} specification and the MP2965 datasheet \removed{, we found the relevant \ac{PMBus} commands}: \Cref{tab:pmbus_cmd} gives the command name, command code, and description of each commands. The first three commands in the table are implemented according to the \ac{PMBus} 1.3 specification~\cite{pmbus13}, while the rest are manufacturer-specific.

\begin{table*}[h]

\begin{center}
\begin{tabular}{ l c l }
 \toprule
 Command name 				& Command code 			& Usage \\
 \midrule
 \texttt{CMD\_PAGE} 	    & \texttt{0x00} 			& Switch between different voltage rails \\
 \texttt{CMD\_OPERATION} 	& \texttt{0x01} 			& \ac{PMBus} override \\
 \texttt{VOUT\_COMMAND} 	& \texttt{0x21} 			& Output voltage settings \\  
 \texttt{READ\_VOUT} 		& \texttt{0x8B} 			& Voltage reading from sensor \\ 
 \texttt{MFR\_VR\_CONFIG} 	& \texttt{0xE4} 			& Enable overclock mode \\
 \texttt{MFR\_OCP\_TOTAL\_SET} 	& \texttt{0xEE} 			& Over-current protection configuration \\
 \bottomrule
\end{tabular}
\end{center}
\caption{Discovered \ac{PMBus} commands on \tSuper.}
\label{tab:pmbus_cmd}
\end{table*}

With \texttt{CMD\_OPERATION}, we can configure the operation mode of the \ac{VRM}. By setting bit 1 of this register, we can enable the \ac{PMBus} override mode. In this mode, the voltage configured in the \texttt{VOUT\_COMMAND} register will override the voltage configuration from the \ac{SVID} bus. Another command that is useful for \pmattackname is \texttt{READ\_VOUT}, as it allows us to read the current voltage of the CPU and establish a baseline for undervolting. 
The \texttt{MFR\_VR\_CONFIG} register is manufacturer-specific. By setting bit 3 or bit 10 and configuring \texttt{CMD\_OPERATION}, we could enable the tracking or fixed voltage overclocking mode, respectively. Bit 8 \texttt{VID\_STEP\_SEL} of \texttt{MFR\_VR\_CONFIG} also allow us to use an alternative mode of \ac{SVID}. In this mode, the \ac{VRM} uses 10\,mV \ac{VID} steps instead of the default of 5\,mV. This makes overvolting up to 3\,V possible, which is well beyond the operating voltage range of the E3-1220 V6 Intel CPU, with a maximum voltage of 1.52\,V~\cite{intel7thgenserver}. We also discovered that the \ac{VRM} has an \ac{OCP} circuit, which can be configured or disabled by another manufacturer-specific register (\texttt{MFR\_OCP\_TOTAL\_SET}). Some \ac{VRM} also support multiple voltage output rails. \texttt{CMD\_PAGE} command is used to select the target rail to send the commands to. 

With these discovered commands, we can now control the CPU voltage through the \ac{PMBus}. In \Cref{sec:pmbus_volt_ctrl}, we detail how this interface is used as part of attack chains for undervolting and overvolting attacks.

\subsection{Jumper Settings}
On the \tSuperMobo motherboard, there are several jumpers that control different functionalities, \changed{including the connection of the \ac{VRM} to other parts of the system}. We kept \changed{all jumpers} in the default status as delivered by the vendor. To avoid confusion, we still list the jumper settings in \Cref{tab:jumper}.
During inspection of the jumper settings, we discovered that the \texttt{SMBDAT\_VRM} and \texttt{SMBCLK\_VRM} jumpers are neither mentioned in the user manual~\cite{supermicro-manual} nor in the quick reference guide~\cite{supermicro-quickref}. Using an oscilloscope while sending \ac{PMBus} commands, we found that these two jumpers can be used for (dis)connecting the \ac{VRM} from/to the \ac{PMBus}. The experiments and attacks described in this paper are conducted under the ``connected'' setting of both jumpers, which according to Supermicro is the default. 

We also found server motherboard without such jumpers, \eg Supermicro X11SPG-TF and \tAsRockServer. For those, the \ac{VRM} is always connected to the \ac{BMC}. We detail our finding on other motherboards in \Cref{sec:eval_other}. It is worth mentioning that to the best of our knowledge, \ac{SGX} attestation does not have the functionality to include the configuration of these (external) jumpers.

\begin{table*}[ht]

\begin{center}
\begin{tabular}{ c l }
 \toprule
 Jumper name 		 		& Description \\
 \midrule

 \texttt{JPME2} 		& Manufacturer mode normal (Default) \\ 
 \texttt{JPB1} 				& BMC enabled (Default) \\  
 \texttt{SMBDAT\_VRM}    & Connect \ac{VRM} data line to \ac{PMBus} \\
 \texttt{SMBCLK\_VRM}     & Connect \ac{VRM} clock line to \ac{PMBus}\\
 \bottomrule
\end{tabular}
\end{center}
\caption{Jumper settings on \tSuperMobo.}
\label{tab:jumper}
\end{table*}

\section{Supermicro's \acs{BMC} and Server Management Interface}\label{sec:BMC}
\changed{Having understood the basic \ac{PMBus} protocol and commands, we next look at different ways to gain access to the \ac{PMBus} and send commands to the \ac{VRM}.}
\changed{To achieve that,} an attacker needs access to the \ac{SMBus}. As described in \Cref{sec:expsetup}, on \tSuper, one of the devices on the \ac{SMBus} is the ASPEED AST2400 \ac{BMC} controller. In this section, we \removed{will} introduce the functionalities and vulnerabilities in these management interfaces that allow us to achieve our main goal---to take control of the \ac{SMBus}.

During the initial investigation of the \ac{BMC}, we found there are mainly three services available: there is a web service running on port 80 (HTTP) and 443 (HTTPS), an \ac{IPMI} over LAN service on port 623, and the SSH service on port 22. Besides, we also found that the \ac{IPMI} service can be accessed through the \ac{KCS} interface from the CPU.

\changed{Some of these interfaces require authentication: to use HTTP, HTTPS, SSH, and \ac{IPMI} -over-LAN, all exposed through Ethernet, one has to authenticate to the \ac{BMC}. The used credentials in this authentication process are individual for each Supermicro motherboard. However, the \ac{IPMI}-over-\ac{KCS} interface does not require any authentication to the \ac{BMC}. Instead, having root privileges on the host OS running on the \ac{CPU} is sufficient to access this interface. One can also use the \ac{IPMI}-over-\ac{KCS} interface to add/remove/modify \ac{BMC} credentials to subsequently login to the Ethernet-exposed interfaces.}

\subsection{SSH Shell}\label{sec:SSH}
Since SSH is one of the most common interfaces that allows us to get a shell and possibly take over the system, we first started our investigation with it. However, the SSH service on \tSuper provides a custom shell called ``ATEN SMASH-CLP System Management Shell''. It only provides limited commands that enable server monitoring and basic management.
Previously, a vulnerability was reported in \cite{ipmi_ssh_root}:  the command \code{shell sh} allows gaining root access from this shell, however, this command was not available on our system-under-investigation. 

\subsection{\acs{BMC} Firmware Analysis}\label{sec:fw_analysis}
\begin{figure}[h]
\centering
\includegraphics[width=0.5\columnwidth]{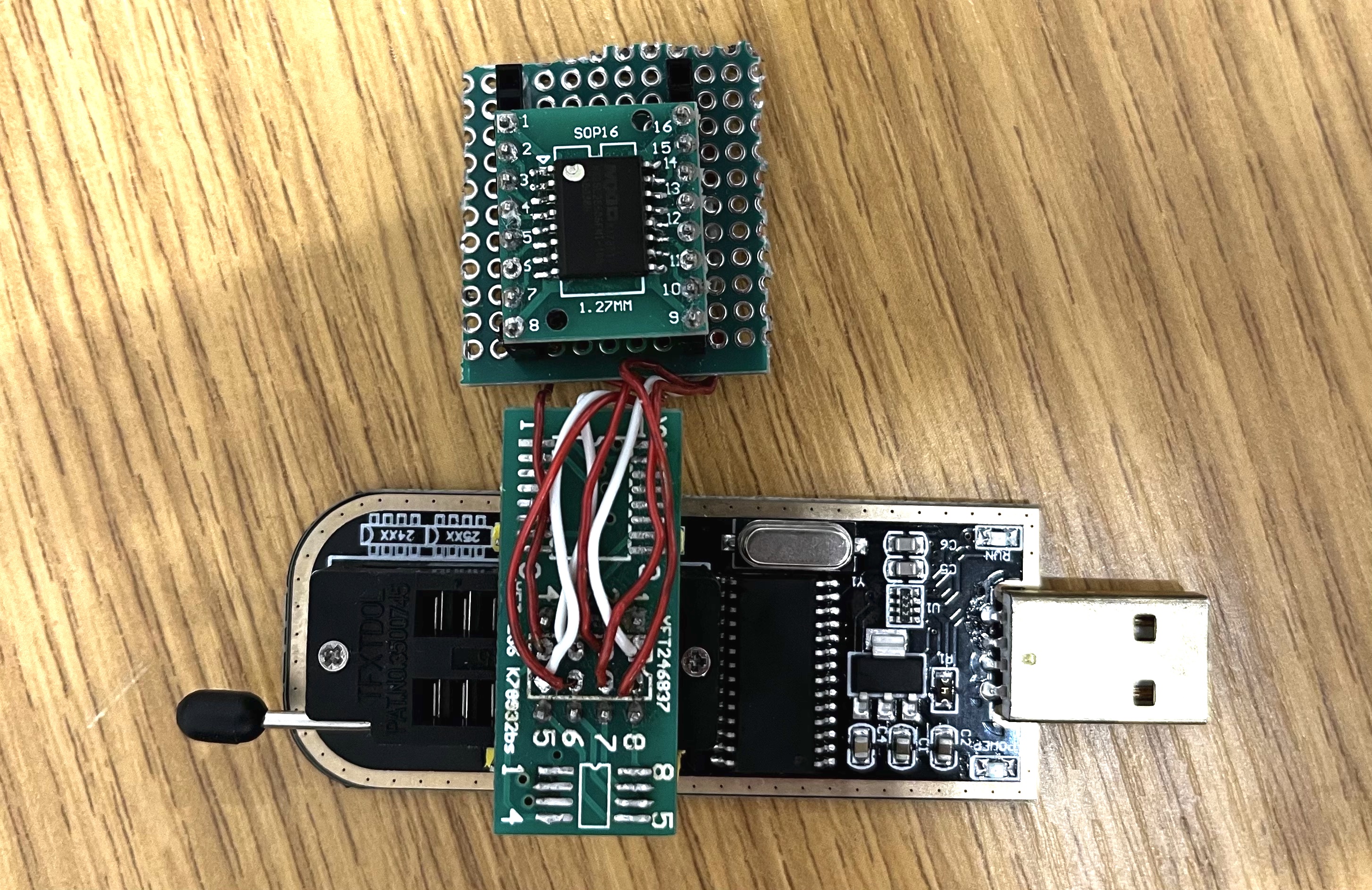}
\caption{Dumping \ac{BMC} firmware with a flash programmer.}\label{fig:fw_dumpping}
\end{figure}

To further investigate the services running on the \ac{BMC} and check if it is possible to enable an SSH root shell, we dumped the firmware of the \ac{BMC} with a CH341A SPI flash programmer as shown in \Cref{fig:fw_dumpping}. This procedure is only used once to assist our analysis, and is not necessary to execute the actual attack. 

We found that the firmware stored in the SPI flash is neither encrypted nor signed. There are five partitions in the firmware, where the second one contains a Linux operating system. The SMASH shell is provided by \texttt{/SMASH/msh} and it is possible to change it to a different shell  by replacing this file. 

The Linux operating system also has an \ac{I2C} kernel module installed, which provides an interface to communicate with the \ac{SMBus}. However, during our testing in \Cref{sec:pmbus_volt_ctrl}, we found that the API provided by this kernel module is not compatible with the commonly used \texttt{libi2c} in \texttt{i2c-tool}\footnote{\url{https://git.kernel.org/pub/scm/utils/i2c-tools/i2c-tools.git/}}. As the result, in \Cref{sec:I2Cdriver}, we opted to write a custom library to use the \ac{I2C} interface of the \ac{BMC} and communicate with the \ac{VRM}. 
 
\subsection{Firmware Upgrade}\label{sec:fwupgrade}

After analysing the firmware, we conclude that it is possible to enable an SSH shell by modifying the firmware. We then started to look for software methods to re-flash the \ac{BMC} \ac{SPI} flash chip. We found that the firmware upgrade functionality of the \ac{BMC} provides a way to do this. There are two interfaces for firmware upgrade: one is through the web interface, the other through the \ac{KCS} interface.

\paragraph{Through Web Interface}
The web interface has a firmware upgrade page that can switch the \ac{BMC} into upgrade mode and allows the user to upload a \ac{BMC} firmware update package. To prevents unauthorised user from upgrading the firmware, there is a login portal. The user is authenticated by the \ac{BMC}. As the \ac{BMC} is a system independent from the OS running on the CPU, users do not need to have privileged access to the OS to be able to use this method. Besides, this web interface can be accessed remotely through Ethernet. 
The remote \ac{BMC} firmware upgrade attack chain described in \Cref{sec:attack_chains} uses this method to upgrade the firmware.

\paragraph{Through \acs{IPMI}-over-\acs{KCS} Interface}\label{sec:ksc_upgrade}

\changed{Crucially, the \ac{BMC} firmware can also be updated through the \ac{KCS} interface, using the following command: \texttt{AlUpdate -f firmware.bin -i kcs -r y}.} \changed{As mentioned,} the \ac{KCS} interface can\removed{only} be accessed from the OS running on the CPU\removed{When using this interface to communicate with the \ac{BMC}, it does not require prior authentication. The only requirement is that the user has}\changed{, only requiring} root access to the \changed{OS, \emph{but not the \ac{BMC} credentials}.} \removed{operating system running on the CPU} \removed{The command for upgrading the firmware via the \ac{KCS} interface is \texttt{AlUpdate -f firmware.bin -i kcs -r y}.}

\paragraph{Firmware Upgrade Package}

\begin{figure}[h]
  \centering
  
  \includegraphics[width=0.9\columnwidth]{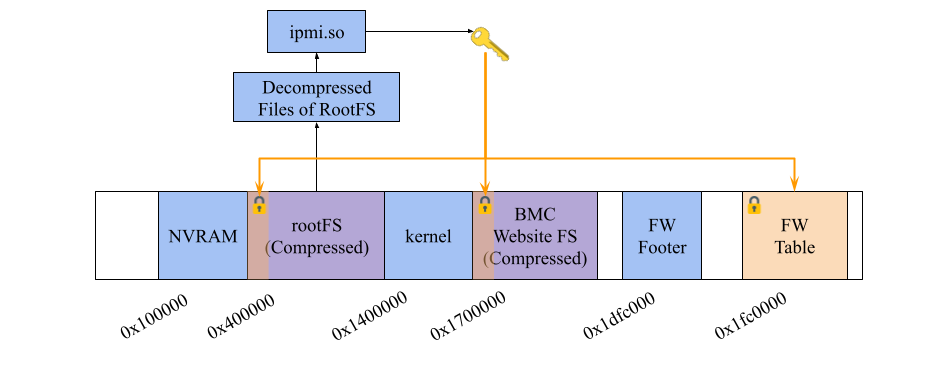}

  \caption{Layout of the \ac{BMC} firmware  upgrade package. The {NVRAM} region stores the current configuration of the \ac{BMC}, the {rootFS} is a LZMA-compressed cramFS file system with only its header encrypted. The {kernel} region stores a Linux kernel image, while the {BMC website FS} is another compressed file system with only the file system header encrypted. The FW Footer starts with a magic value \texttt{ATENs\_FW} and contain information about the firmware version, checksum, etc. The {FW Table} is an encrypted region and stores a table of the image layout. All encrypted region of the firmware can be decrypted with a key extracted from \textit{ipmi.so} on the \textit{rootFS}. \label{fig:bmc_fw_layout}}
\end{figure}

After finding the firmware upgrade interface, the next step is to produce an upgrade package that can be uploaded to the \ac{BMC}. We started with the analysis of the structure of the upgrade package. 
\Cref{fig:bmc_fw_layout} shows the layout of a firmware upgrade package. Previous work by \cite{eclypsium_2018} founds that in the firmware upgrade package, there is a region that contains a magic value (\texttt{ATENs\_FW}), a half-length CRC checksum, and the length of each section. We call this part the firmware footer. There is also a region containing metadata of the firmware image, including the name of each region and their length and CRC, starting with \changed{``\texttt{[img]}''}. We refer to this region as firmware table.  In the X11 series, the firmware table, the \textit{file system header} of the root file system and the website \textit{files system header} are AES-CBC encrypted. However, the files in these regions are not encrypted, but only LZMA compressed. As a result, the key of the AES-CBC encryption can be recovered from the \texttt{ipmi.so} file on the root file system. 

With this information, we can modify the firmware and construct a valid firmware upgrade package for the web interface. We discuss firmware repacking in detail in \Cref{sec:firmware_repack}.

\subsection{IPMI \acs{I2C} functionality}
When exploring the functionalities of \ac{IPMI}, we \changed{also} found that the interface also allows direct sending \ac{I2C} packets with the \texttt{ipmitool i2c} command. This can be used either through the Ethernet or \ac{KCS} \ac{IPMI} channel. The authentication requirement for using \ac{IPMI}-controlled \ac{I2C} is the same as those described in \Cref{sec:fwupgrade}. As shown in \Cref{sec:takeover_ipmi}, we can use this functionality for direct access to the \ac{SMBus}/\ac{PMBus} \textit{without} modifying \ac{BMC} firmware. 

\section{Practical Experiments}
\label{sec:practical}
\begin{figure}[h]
  \centering

  \resizebox{0.6\columnwidth}{!}{
    \def\svgwidth{1\textwidth}
    \footnotesize
    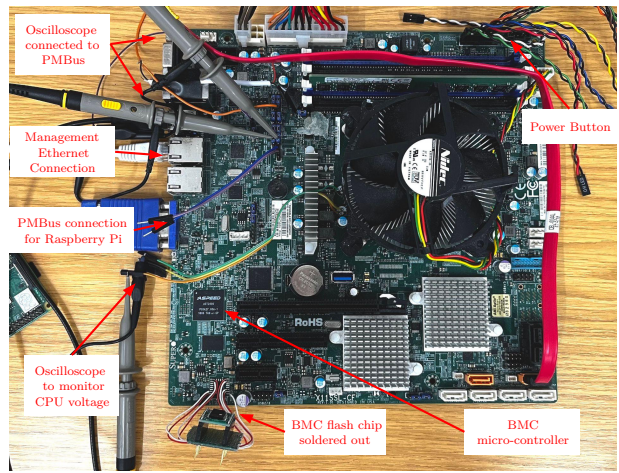
  }

  \caption{Setup of the \tSuper for practical experiments. These connections are for experiments only; physical access is not required in the actual attack. \label{fig:supermicro_motherboard}}
\end{figure}

\removed{In this section,} \changed{Finally, using the results from the previous sections,} we explain how \changed{to construct practical \ac{PoC} attacks for \pmattackname.}\removed{we discovered the system configuration and build a \ac{PoC} of \pmattackname.} Some of our experiments require physical access to the system to understand the hardware configuration (with an overview shown in \Cref{fig:supermicro_motherboard}). Note however that physical access is not required when performing \pmattackname attacks on a real-world system, as the hardware components and connections are identical for a given motherboard model.

\subsection{PMBus-based Voltage Control}\label{sec:pmbus_volt_ctrl}
To understand the configuration and capabilities of using the \ac{PMBus} to control the CPU voltage, we conducted two experiments. Firstly, we used the ``probe and verify'' method to find the \ac{I2C} address of the \ac{VRM}. Then we tried different ways of sending commands to \ac{VRM} to change the voltage.

\paragraph{Discovering the \ac{VRM} Address}\label{sec:discover_vrm_addr}
\changed{Finding the \ac{I2C} address of the \ac{VRM} is the first step of \pmattackname.}
The easiest way to explore the \ac{I2C} buses is to use the interface provided by the OS. There are two \ac{I2C} buses that can be used from the \ac{OS} running on the CPU: \texttt{i2c-0} is shown by default, while \texttt{i2c-1} requires the \texttt{i2c\_i801} kernel module to be loaded. To find all available devices on both \ac{I2C} buses, we ran the \texttt{i2cdetect} tool on them. We found that there are 12 devices in total connected to the \ac{I2C} bus. The full list of device addresses can be found in \Cref{appdx:i2cdetect}.

To \changed{then} determine which device is a \ac{VRM}, we use the result of the standard \ac{PMBus} command, \texttt{READ\_VOUT}, as an indicator. The Plundervolt\ \cite{murdock2019plundervolt} attack showed that the normal operating voltage of the CPU should be greater than 0.55\,V, thus, if the voltage read by \texttt{READ\_VOUT} is within this range, it may be a \ac{VRM}. Of the 12 devices detected, only one device with address \texttt{0x20} on \ac{I2C} bus 1 responded with a value in this voltage range. We hence suspect this device is the \ac{VRM}. 
To verify the result, we also used \texttt{MFR\_ADDR\_PMBUS} (\code{0xE1}) command found in the {MP2965} datasheet\ \cite{mp2965} to read the \ac{PMBus} address of the device. The result is \texttt{0x20}, which confirms our finding.  

\paragraph{Changing CPU Voltage with \acs{PMBus} Commands}\label{sec:pmbus_cmd_seq}
\changed{Having identified the \ac{VRM}, one can next attempt to send commands to change the CPU voltage.}

\begin{figure}[ht]
  \centering

  \resizebox{0.9\columnwidth}{!}{
    \def\svgwidth{1.2\textwidth}
    \footnotesize
\begingroup%
  \makeatletter%
  \providecommand\color[2][]{%
    \errmessage{(Inkscape) Color is used for the text in Inkscape, but the package 'color.sty' is not loaded}%
    \renewcommand\color[2][]{}%
  }%
  \providecommand\transparent[1]{%
    \errmessage{(Inkscape) Transparency is used (non-zero) for the text in Inkscape, but the package 'transparent.sty' is not loaded}%
    \renewcommand\transparent[1]{}%
  }%
  \providecommand\rotatebox[2]{#2}%
  \newcommand*\fsize{\dimexpr\f@size pt\relax}%
  \newcommand*\lineheight[1]{\fontsize{\fsize}{#1\fsize}\selectfont}%
  \ifx\svgwidth\undefined%
    \setlength{\unitlength}{755.00001526bp}%
    \ifx\svgscale\undefined%
      \relax%
    \else%
      \setlength{\unitlength}{\unitlength * \real{\svgscale}}%
    \fi%
  \else%
    \setlength{\unitlength}{\svgwidth}%
  \fi%
  \global\let\svgwidth\undefined%
  \global\let\svgscale\undefined%
  \makeatother%
  \begin{picture}(1,0.09933775)%
    \lineheight{1}%
    \setlength\tabcolsep{0pt}%
    \put(0,0){\includegraphics[width=\unitlength,page=1]{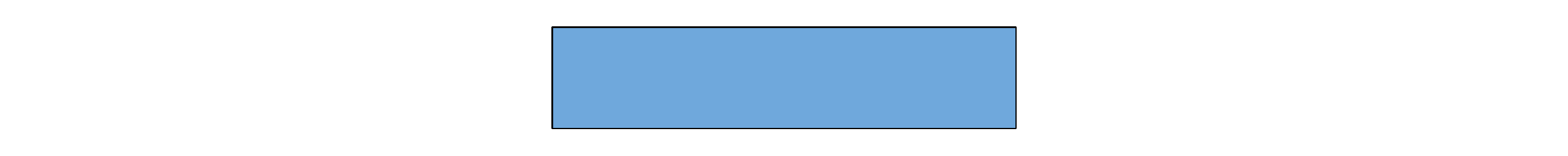}}%
    \put(0.41722907,0.05372195){\color[rgb]{0,0,0}\makebox(0,0)[lt]{\lineheight{1.25}\smash{\begin{tabular}[t]{l}Set target voltage to \end{tabular}}}}%
    \put(0.41779464,0.03186769){\color[rgb]{0,0,0}\makebox(0,0)[lt]{\lineheight{1.25}\smash{\begin{tabular}[t]{l}VOUT\_COMMAND \end{tabular}}}}%
    \put(0,0){\includegraphics[width=\unitlength,page=2]{PMBusCommands.pdf}}%
    \put(0.03766657,0.05372195){\color[rgb]{0,0,0}\makebox(0,0)[lt]{\lineheight{1.25}\smash{\begin{tabular}[t]{l}Configure VOUT\_OPERATION \end{tabular}}}}%
    \put(0.0549953,0.03186769){\color[rgb]{0,0,0}\makebox(0,0)[lt]{\lineheight{1.25}\smash{\begin{tabular}[t]{l}with PMBus Override Mode\end{tabular}}}}%
    \put(0,0){\includegraphics[width=\unitlength,page=3]{PMBusCommands.pdf}}%
    \put(0.70021944,0.04279482){\color[rgb]{0,0,0}\makebox(0,0)[lt]{\lineheight{1.25}\smash{\begin{tabular}[t]{l}Set Bit 3 of MFR\_VR\_CONFIG \end{tabular}}}}%
    \put(0,0){\includegraphics[width=\unitlength,page=4]{PMBusCommands.pdf}}%
  \end{picture}%
\endgroup%

  }

  \caption{Command sequence to change the voltage via \ac{PMBus}. \label{fig:pmbus_cmd_seq}}
\end{figure}
In the datasheet of the MP2965~\cite{mp2965}, we found an ``overclocking'' procedure that can be used for \changed{this purpose}\removed{configuring the CPU voltage}. There are two overclocking modes, \textit{tracking mode} and \textit{fix mode}. In \pmattackname, we mainly use the {fix mode} to set a defined voltage. 

In the fix overclocking mode, the \ac{VRM} uses the \ac{VID} configured with the \ac{PMBus} command \texttt{VOUT\_COMMAND} and ignores the configuration from the \ac{SVID} bus. \Cref{fig:pmbus_cmd_seq} shows the steps of using this mode to change voltage. First, we need to configure two registers: The first one is \texttt{VOUT\_OPERATION}; by setting the first bit of this register, we enable \ac{PMBus} override mode. We also have to set bit 3 of \texttt{MFR\_VR\_CONFIG} to make the \ac{VRM} act on these changes. After this, the voltage supplied to the CPU will be changed according to the configuration in \texttt{VOUT\_COMMAND}. 
To send this \ac{PMBus} command sequence and change the CPU voltage, we wrote a \ac{PoC} with the \texttt{libi2c}. This \ac{PoC} can be compiled and run under Linux.

\paragraph{``Stalls'' caused by PMBus Commands}
The experiments in \Cref{sec:discover_vrm_addr} also show that the \ac{VRM} responds to the \ac{PMBus} commands sent from the CPU. One may thus assume that it would then be straightforward to directly send \ac{PMBus} commands to change the CPU voltage with this method. However, we found that the CPU  stalls after sending the \texttt{MFR\_VR\_CONFIG} command to actually configure the \ac{VRM} to use the new voltage. This will make the CPU voltage being kept at the changed value with no way to change it back. 
This phenomenon raised two questions: Is the CPU stall caused by a crash or a recoverable halt? If it is caused by a recoverable halt, will this protect against targeted undervolting fault injection? 

To answer this, we connected a Raspberry Pi to the \ac{PMBus} to directly control the \ac{VRM}. The \ac{I2C} interface to the \ac{VRM} is exposed with two pins, \texttt{SDA} and \texttt{SCL}. As shown in \Cref{fig:supermicro_motherboard}, we connected the \ac{I2C} interface of the Raspberry Pi to these pins. 

In the first experiment, we sent a command to disable overclocking after the stall happens. It appears that with the \ac{VRM} reconfigured to normal mode, the CPU recovers from the stall situation if the undervolting value is not too low. This shows that the stall is caused by a recoverable halt and not a crash. 
The second experiment is used to find out if the halt will prevent the fault from happening. In this experiment, we used the {CRT-RSA} \ac{PoC} of the Plundervolt attack. With the CPU running this \ac{PoC}, we used Raspberry Pi to send \ac{PMBus} commands to produce voltage glitches. We found that with glitches with gradually lower voltage, an exploitable fault happens with the CRT-RSA calculation. 

Hence, in summary, the ``stall'' phenomenon will prevent the \ac{PMBus} attack from being conducted by the CPU-\ac{VRM} \ac{I2C} interface, but it does not prevent the fault caused by undervolting from having an impact on CPU calculations. 

\paragraph{Voltage Control with BMC}\label{sec:I2Cdriver}
Because our attempt of voltage glitching failed with the \ac{PoC} running on the CPU, we started to look into the \ac{BMC}-\ac{VRM} \ac{I2C} interface. In the \ac{BMC} firmware dumped in \Cref{sec:SSH}, we found the \texttt{i2c.ko} kernel module, which provides a driver for the \ac{I2C} interface. However, this module does not implement a standard \removed{(and required)} \texttt{ioctl()} for \ac{I2C} devices\changed{, which is required for using \texttt{libi2c}}. This means that the \changed{above} \ac{PoC}\removed{in Section~\ref{sec:pmbus_cmd_seq}}, which uses \changed{this standard \ac{I2C} library}, cannot be used to communicate with this kernel module.

As the kernel module in the firmware did not implement the standard \ac{I2C} API, we had to find another way to utilize the \ac{BMC}'s \ac{I2C} interface. With the help of the \ac{I2C} driver in the latest Linux kernel~\cite{ast2400-i2c-devtree,ast2400-i2c-kdriver}, we found that there are 14 \ac{I2C} \changed{interfaces} on the AST2400 \ac{BMC} controller. Each has a set of memory-mapped registers to control the interface. We also found the setup and message sending/receiving sequence of the \ac{I2C} interface. 
We then created a small library to directly write these registers and send \ac{I2C} bus commands from the \ac{BMC} CPU to the address of the \ac{VRM}. By monitoring the \ac{I2C} activity with an oscilloscope (this was only required for debugging and during development), we found that the \ac{I2C} bus 2 (counted from bus 0) of the \ac{BMC} has the \ac{VRM} connected.

\subsection{Enabling SSH Access and Firmware Repacking}\label{sec:firmware_repack}
\changed{Modification of the firmware can be used to obtain a root shell on the \ac{BMC}.}
With the ``Supermicro BMC firmware image decryptor''~\cite{smcbmc-tool} and a modified version of the ``ipmi firmware tool''~\cite{ipmi-fw-tool} with added support for X11 images, we were able to extract the firmware encryption key and decrypt the file system header. With these, we can unpack and modify the full root file system. 

\removed{The main purpose of this firmware modification is to obtain a root shell on the \ac{BMC}.} As described in \Cref{sec:fw_analysis}, \texttt{/SMASH/msh} provides the shell for SSH service. To enable full \changed{root} shell access, we replaced this file with a shell script with a single line to execute \texttt{/bin/sh}. Besides, as the SSH service is running with root privileges, with the shell redirected to \texttt{sh}, we could obtain a root shell once connected to the SSH. 

To repack the image, we modified the ``Supermicro BMC firmware image decryptor'' tool to add firmware encryption support and constructed a firmware package with a valid footer and firmware table. We successfully tested and installed this modified firmware package both with the web firmware upgrade interface and the IPMI firmware upgrade interface via the \texttt{AlUpdate} tool.

\subsection{Attack Chains for \ac{PMBus} Access}\label{sec:attack_chains}

In this section, we discuss three possible attack chains to take over the \ac{PMBus} with the techniques shown in the previous sections. The attacker can use any of these attack chains and change the CPU voltage to perform \pmattackname attacks, \ie to over/undervolt the CPU.

\paragraph{Remote \ac{BMC} Firmware Upgrade}
The first attack chain assumes a malicious insider threat model. This attack chain makes use of the web or IPMI interface through the \ac{BMC} Ethernet connection. To use this interface, the attacker needs to have access to the \ac{BMC} management Ethernet port or the shared management Ethernet port \code{eth0} on the system. Besides, the attacker needs to obtain valid credentials to login to the \ac{BMC}. 

In detail, the attacker can first use the method described in \Cref{sec:firmware_repack} to repack the \texttt{SMT\_X11\_163} firmware upgrade package from~\cite{bmc-fw} to enable SSH \code{root} access to the \ac{BMC}. Then, they can upload the firmware with the web management interface or the \ac{IPMI} management interface over Ethernet. With the SSH interface enabled, the attacker can cross-compile the voltage-changing \ac{PoC} described in \Cref{sec:I2Cdriver} for the \ac{BMC}, and then upload and execute it to send \ac{PMBus} commands. We used \texttt{base64 -d > /tmp/i2c-pmbus-send} to upload our exploit code due to the unavailability of the SCP service on the \ac{BMC} OS.

\paragraph{Local \ac{BMC} Firmware Upgrade}
Similar to the first, this attack chain also involves a firmware upgrade for code execution on the \ac{BMC}. However, we use the \ac{KCS} interface discussed in \Cref{sec:ksc_upgrade} to upgrade the firmware. The attacker does not require access to the management Ethernet plane, instead, only \code{root} privileges on the OS running on the CPU is required. This is \eg relevant for data centers that host bare metal machines for customers or for malware/ransomware that has obtained \code{root} through other exploits.

\paragraph{\ac{IPMI} Interface}\label{sec:takeover_ipmi}
The third attack chain uses the \ac{IPMI} \ac{I2C} functionality. An attacker with \code{root} access on the CPU OS or access to the management port of the \ac{BMC} can use this interface to send commands to any \ac{I2C} device that is connected to the \ac{BMC}. 
The command used for sending the raw \ac{I2C} packets is shown in \Cref{lst:ipmi_cmd}. The \ac{I2C} mapping of this interface is the same as found during the initial investigation in \Cref{sec:I2Cdriver}. The \ac{VRM} is at address \texttt{0x20} on bus 2. However, since the last bit of the first packet of \ac{I2C} indicates the type of operation (read or write), we need to shift the device address left by one bit and set the last bit accordingly when using this interface to control \ac{PMBus}.

\begin{lstlisting}[language=bash, label={lst:ipmi_cmd}, caption={\ac{IPMI} command for sending \ac{I2C} packets.}, captionpos=b]
ipmitool i2c bus=2 0x40 <PMBus Command> <PMBus Data>
\end{lstlisting}

\section{Undervolting and Overvolting Attacks}
\label{sec:attacks}
\begin{figure}[ht]
  \centering
  
  \resizebox{0.95\columnwidth}{!}{
    \def\svgwidth{1.5\textwidth}
    \footnotesize
    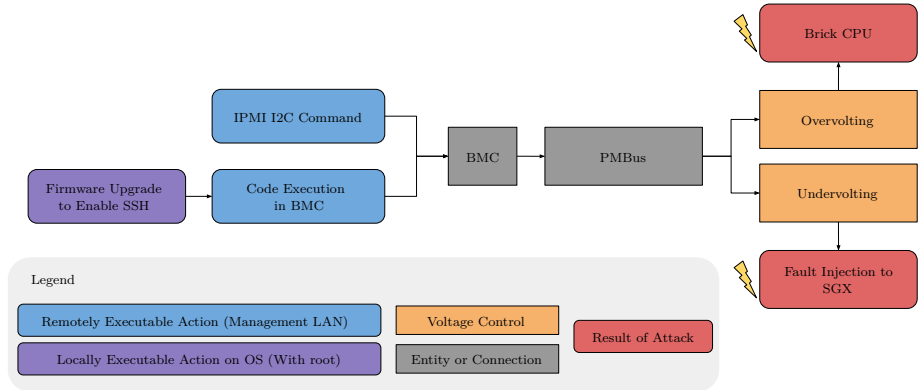
  }

  \caption{Overview of the \pmattackname attack. With root access to the \ac{OS} or access to the \ac{BMC} via Ethernet or \ac{KCS}, the attacker can perform a malicious firmware upgrade of the \ac{BMC} and then takeover the \ac{PMBus}. The attacker can also use the \code{ipmi i2c} command to directly control the \ac{PMBus} via \ac{BMC}. With control over the \ac{CPU} voltage, the attacker can overvolt to brick the CPU or undervolt to inject faults into \ac{SGX}. \label{fig:attack}}
\end{figure}

In this section, we show how under/overvolting through the \ac{PMBus} \removed{(\eg via a compromised \ac{BMC})} leads to attacks on \ac{SGX} and also permanent physical damage to the CPU\removed{, \eg by malicious software}. \changed{The attack requires any flaw that gives a software attacker access to the \ac{PMBus}. As mentioned in \Cref{sec:attack_chains}, this can \eg be a malicious firmware upgrade or the use of the \ac{IPMI}-to-\ac{I2C} functionality. The attack is generic in the sense that \emph{various} flaws can lead to the same outcome: remote fault injection attacks on \ac{SGX} and bricking the CPU.} \Cref{fig:attack} shows an overview of the attacks. 

\subsection{Undervolting Attack against Intel \acs{SGX}}
\paragraph{Adversary Model}
As mentioned in \Cref{sec:adversary}, we assume \removed{the standard SGX}\changed{a} threat \changed{model}\removed{where an attacker has full \code{root} access to the system, but no (or limited) physical access. We also consider an attack involving a malicious insider, as \ac{SGX} is often assumed to provide a \ac{TEE} that protects against \eg a cloud provider.} \changed{where an attacker (including a malicious insider) has full software access to the system but no (or limited) physical access. More precisely, the attacker has \code{root} access to the OS and software access to the \ac{BMC} via the \ac{KCS} interface or Ethernet.}
All attack chains described in \Cref{sec:attack_chains} can generally be used under this threat model. It is worth mentioning that the attack that uses \code{ipmitool} through the \ac{KCS} interface does not require knowledge of the \ac{BMC} credentials. A privileged local user on a compromised host CPU can thus use \code{ipmitool}  to inject fault into \ac{SGX} purely from software.

\paragraph{Proof of Concept}

We used the same PoC code as Plundervolt/VoltPillager~\cite{murdock2019plundervolt}. Before injecting the voltage glitch, we use the attack chain described in \Cref{sec:attack_chains} to gain control of the \ac{PMBus}. 

To start with, we used the multiply operation as the first target, as it is a simple target to fault. By gradually lowering the CPU voltage with the \ac{PMBus} commands sent by the \ac{BMC} while running the Plundervolt/VoltPillager PoC on the CPU, we successfully injected faults into the multiply operation (in our experiments at voltage 0.845\,V with the CPU running at 2\,GHz. 

To verify the fault injection also works for encryption operations running in \ac{SGX}, we ran the CRT-RSA signature \ac{PoC} from Plundervolt/VoltPillager, with an RSA signature computed inside an enclave using the \removed{ipps functions} \changed{Intel Integrated Performance Primitives (Intel IPP) cryptography library functions~\cite{ipps-rsa-dev-ref}}. Again, we could obtain faulty signatures as shown in Listing~\ref{lst:faults}. Furthermore, we confirmed that these faulty values could be used to factor the RSA modulus and recover the private RSA key using the Lenstra attack~\cite{BonehDemilloLipton97}.

\begin{mdframed}[backgroundcolor=black!5,leftmargin=0.5cm,hidealllines=true,%
  innerleftmargin=0.2cm,innerrightmargin=0.2cm,innertopmargin=0.0cm,innerbottommargin=-0.7cm ]
\begin{lstlisting}[basicstyle=\scriptsize\ttfamily,breakatwhitespace=true, breaklines=true, label={lst:faults}, caption=Faulty CRT-RSA decryptions/signatures generated by the respective ipps functions., captionpos=b]
// Faulty calculation 1
0x3f, 0xe0, 0xb8, 0x74, 0x04, 0x18, 0x9c, 0xed, 0x91, 0x1a, 0x02, 0x12, 0x2a, 
0xce, 0x89, 0xf8, 0x32, 0x00, 0xdc, 0x05, 0x15, 0x53, 0x72, 0x8d, 0x84, 0x00, 
0xd3, 0x67, 0xbe, 0xa1, 0xc2, 0x40, 0x76, 0xbc, 0x8c, 0xd8, 0xfe, 0xb1, 0x00, 
0xd7, 0x9e, 0x0e, 0xb6, 0xac, 0x61, 0xc0, 0xec, 0x9c, 0xf7, 0x7e, 0xbc, 0x4b, 
0xde, 0x18, 0xa5, 0xa4, 0x1c, 0x74, 0xc4, 0xb5, 0x6a, 0x8d, 0xd3, 0xb1, 0x35, 
0xf9, 0xad, 0x0b, 0xe3, 0x4a, 0x01, 0x52, 0xd4, 0xc6, 0xb2, 0x95, 0xbc, 0xdc, 
0xad, 0x61, 0x8e, 0x07, 0x84, 0x4d, 0xe3, 0xa7, 0xff, 0xf0, 0xd1, 0xa0, 0xd4, 
0x58, 0x9f, 0xbc, 0x37, 0x0b, 0xa8, 0x91, 0x83, 0x15, 0x7b, 0xee, 0x28, 0x83, 
0x12, 0x4a, 0x89, 0x61, 0x1e, 0x2c, 0xe1, 0x02, 0x2f, 0x08, 0x4d, 0x5b, 0x04, 
0x92, 0x5e, 0x31, 0xd0, 0x7e, 0x94, 0x85, 0xd0, 0xce, 0x75, 0x4a, 0x00, 0x00, 
0x00, 0x00, 0x00, 0x00, 0x00, 0x00, 0x00, 0x00, 0x00, 0x00, 0x00, 0x00, 0x00,
[... zeroes left out ...]
Incorrect result!
\end{lstlisting}
\end{mdframed}

\paragraph{Reproducibility of CRT-RSA Fault Injection}
\changed{To further evaluate the reproducibility of the attack, we setup an automated testing environment by connecting a Raspberry Pi to an Ethernet port (\texttt{eth0}) and the power button of the motherboard. We ran a Python script to repeat the following steps numerous times:
\begin{enumerate}[leftmargin=*, topsep=1pt,itemsep=-3pt]
    \item Upload the exploit for controlling the CPU voltage to \ac{BMC} via an SSH connection.
    \item SSH into the \ac{OS} running on the host \ac{CPU} and trigger CRT-RSA signing in an \ac{SGX} enclave.
    \item Run the \pmattackname exploit on the \ac{BMC} to gradually lower the CPU voltage while the signature is computed in the \ac{SGX} enclave.
    \item Stop lowering the CPU voltage when a fault occurs.
    \item Record the result and cleanup.
    \item If no faulty result is output, the system may have crashed due to too low voltage. In this case, we use the connection to the motherboard power button to reboot the system and wait  to allow the system to boot into a stable status. 
\end{enumerate}

In total, we conducted 253 tests within 545\,min. Of those, faults occurred in 194 tests. 66 of these faulty results could be used to successfully recover the correct RSA private key using the Lenstra attack, which translates to a success rate of 26\%. On average, a useful fault could be obtained within 9 minutes. 
}

\subsection{Overvolting to Permanently Brick a CPU}
Apart from the undervolting attack to extract keys from an \ac{SGX} enclave, we also discovered another attack, which is an overvolting attack that can permanently destroy the CPU. 

\paragraph{Adversary Model}
In this attack, as described in \Cref{sec:adversary}, we assume an attacker who has \code{root} privilege on the host CPU. For example, this could be in the case that an attacker has placed ransomware on a system and threatens to damage the CPU unless a ransom is paid. Clearly, \code{root} should have full control of all software running on the CPU, but \emph{should not} be able to cause any physical damage to the system. The attack chain described in \Cref{sec:attack_chains} using \texttt{ipmitool} with \ac{KCS} can be used within this threat model. 

\paragraph{Proof of Concept}
To overvolt the \ac{CPU}, we firstly configure the \texttt{MFR\_VR\_CONFIG} register of the \ac{VRM} to use the 10\,mV \ac{SVID} table. This allows changing the CPU voltage up to 3\,V. We also disabled the over-current protection by reconfiguring the \texttt{MFR\_OCP\_TOTAL\_SET} register. Then we used the voltage changing procedure to change the CPU voltage to a value much higher than the normal operating voltage. 

\begin{figure}[h]
       \centering
       \includegraphics[width = 0.7\textwidth ]{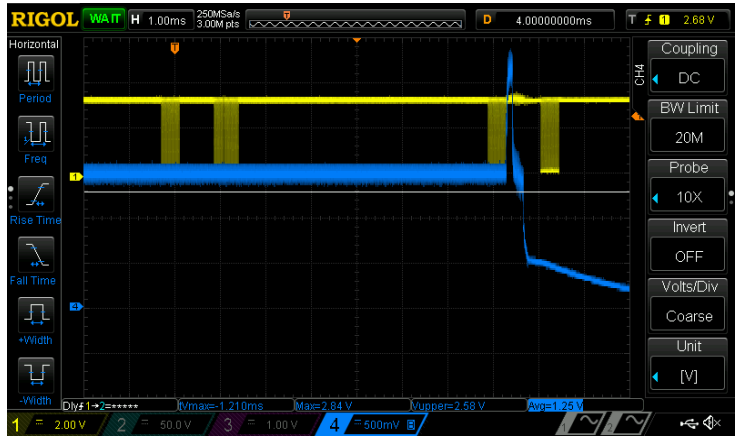}
       \caption{Oscillocope capture of voltage change during overvolting, \texttt{VOUT\_COMMAND} set to \texttt{0xFF} (with 10\,mV VID table). Yellow: PMBus clock, blue: $V_{cpu}$. $V_{cpu}$ shoots up to 2.84\,V during overvolting. }
       \label{fig:osci_voltage_change}
\end{figure}

We found that this procedure allows changing the CPU voltage up to $\sim$2.84\,V \changed{for $\sim$1\,ms}, which is outside the typical operating range of Intel CPUs. By increasing the voltage beyond the specified operating voltage range (0.55\,V--1.52\,V)~\cite{intel7thgenserver} of a 7th Gen Intel E3-1220V6 CPU \changed{two times}, we permanently destroyed the CPU and left the system in an unbootable state \changed{within a few seconds}. We successfully repeated the experiment with a second, identical CPU. An example of overvolting is shown in \Cref{fig:osci_voltage_change}. 

For environmental and financial reasons, we were satisfied after successfully destroying two CPUs and decided to not perform further experiments in that regard.

\section{Evaluation of other Server Motherboards}\label{sec:eval_other}

As we found the \ac{PMBus} to be a common interface present on server motherboard, we decided to investigate other manufacturers as well. To facilitate larger-scale testing of this, we wrote a tool called \pmdetect. With this tool, we scan the system for a \ac{PMBus} connection and try to detect the \ac{VRM} address. We applied this tool to several other systems, including an ASRock rack motherboard (\tAsRockServer) and a \tSuperMoboNew motherboard (kindly provided by Supermicro for testing). We then conducted further analysis of these systems to check if they are vulnerable to any \ac{PMBus}-related attack. 

\paragraph{\pmdetect Tool for \ac{VRM} Detection}
Based on the \ac{VRM} detection process mentioned in \Cref{sec:discover_vrm_addr}, we built the \pmdetect tool to automatically scan all addresses of a specified \ac{I2C} bus for \acp{VRM}. During testing, we found that the implementation of \ac{PMBus} and usage of the \ac{VRM} is different between motherboard, and the most stable command to identify a \ac{VRM} is \texttt{READ\_TEMPERATURE} (\code{0x8d}). We use the response to this command as an initial indicator to identify whether a \ac{VRM} is present, and then use the \ac{VRM} detection process from \Cref{sec:discover_vrm_addr} to verify the result.

Moreover, as the capabilities and voltage changing sequence can differ between \ac{VRM} vendor, we added an additional procedure to detect the vendor of the \ac{VRM}. For this, we use the result of reading \texttt{ISL\_DEVICE\_ID} (\code{0xad}) as an indicator for Intersil \acp{VRM} and \texttt{SVID\_VENDOR\_PRODUCT\_ID} (\code{0xbf}) for MPS, respectively. Detection based on \texttt{ipmi i2c} is also implemented for detecting the connection between \ac{VRM} and the \ac{BMC} as mentioned in \Cref{sec:attack_chains}. An example output of \pmdetect with \tSuperMobo is shown in \Cref{appdx:detect_res}, while \Cref{tbl:tested_mobo} shows a summary of the motherboard tested and the scan result for \acp{VRM} with \pmdetect.
We are aware that our testing---restricted by (lack of) access to server hardware---\removed{to} only gives a very limited picture of the use of \ac{PMBus} and \acp{VRM} on server hardware. We hence decided to open-source \pmdetect and build on community efforts in the future to obtain a better view of the \ac{PMBus} landscape. 

\begin{table}[ht]
\centering
\resizebox{\columnwidth}{!}{
\begin{tabular}{ccccc}
\toprule
Name                & \ac{BMC} & Chipset  & \ac{VRM} Address                        & \ac{PMBus} Connects to           \\
\hline
\tSuperMobo         & AST2400 & C232  &\texttt{0x20}                                & BMC \& CPU          \\
\tSuperMoboNew      & AST2600 & C621A  &\texttt{0x30} \& \texttt{0x34}              & ---                   \\
\tAsRockServer      & AST2500 & C246  &\texttt{0x60}                                & BMC \& CPU          \\      
\bottomrule
\end{tabular}
}
\caption{Tested motherboards and their \ac{VRM} detection result. \label{tbl:tested_mobo}}
\end{table}

\subsection{ASRock Power-Down Attack}
The \tAsRockServer motherboard uses an Intel Xeon E-2124 CPU with an Intel C246 Chipset and ASPEED AST2500 \ac{BMC} \changed{with login credentials defaulting to \texttt{ADMIN:ADMIN}}. We used the \pmdetect tool together with manual probing and found that the \ac{VRM} of this motherboard is connected to both the \ac{BMC} and \ac{I2C} bus of the CPU. In the following attack, we assume that the attacker is a user on a baremetal server with root access in the OS.

The \ac{VRM} used on this motherboard is an ISL69138. Because it is made by a different manufacture compared to the MP2955, the voltage changing \ac{PMBus} command sequence used for the MP2955 does not work with this \ac{VRM}. Due to lack of documentation of this procedure, we at the moment could not precisely overvolt or undervolt the CPU via the \ac{PMBus}. Yet, we discovered a new attack to disable the \ac{VRM} and force power-down the CPU, leaving the system in a (temporary) inoperable state. 

\pmdetect shows that the \ac{VRM} is at address \code{0x60} on \ac{I2C} bus 2 of the host \ac{CPU}. Different to the findings for the \tSuperMobo, this \ac{VRM} uses \ac{PMBus} registers on page \code{0x1} instead of the default \code{0x0}. We then issue the \texttt{ON\_OFF\_CONFIG}  (\code{0x02}) and \texttt{OPERATION} (\code{0x01}) commands: We configure the \texttt{OPERATION} to ``Immediate Off'' and set the ``source of enable'' only to \texttt{ON\_OFF\_CONFIG}. This results in a immediate power-off of the \ac{VRM} and crashes the system. 

During testing, we found the \ac{PMBus} is only writable from the CPU with \ac{IPMI} over \ac{KCS} interface, but not from the \ac{BMC} with \texttt{ipmi i2c} commands. As the result, it is not possible for the administrator of the system to remotely configure the \ac{VRM} back to a normal state. Simply issuing the \texttt{ipmi powercycle} command with \ac{IPMI} over LAN will leave the system in a infinite boot loop. To recover from this attack, the administrator has to physically power-cycle the system, which might increase downtime in a \ac{DoS} scenario.

This shows that \ac{PMBus} as an attack vector does not only affect \tSuperMobo, but also can have impact on servers from other manufacturers. Besides we believe that it might also be possible to conduct \ac{CPU} bricking attacks if the \ac{PMBus} voltage changing sequence of Intersil \ac{VRM} is known. We leave this for future work. 

\subsection{Other Supermicro X11 Motherboards}
We also ran the \pmdetect tool on X11SPG-TF and X11SSE-F Supermicro server motherboards---in both cases, the \ac{VRM} was reachable in the default configuration. To test if they are vulnerable to \pmattackname, we sent \ac{PMBus} commands through \texttt{ipmi i2c} commands and successfully undervolted them to crash the system. This shows that the attack chain through the \ac{IPMI} interface is valid on these systems. As the systems were provided by a third party for remote testing, we were not able to attempt overvolting and similar, destructive experiments, but believe these motherboards to be equally affected.

\subsection{Supermicro X12 Motherboards}
We disclosed the vulnerability to Supermicro in May 2022. They confirmed the issue and also provided a X12 generation Server for further testing. This system, \tSuperMoboNew, features a dual Intel Xeon Gold 6330 CPU, Intel C621A Chipset, and AST2600 \ac{BMC}.
Our investigation \removed{on this motherboard} shows that mitigations \changed{has already} been implemented on this motherboard to break the attack chain of \pmattackname\ \changed{before we reported the attack to Supermicro}. Firstly, the firmware upgrade package is properly signed with RSA and verified during the firmware upgrade process, which prevents malicious firmware \changed{uploads to} the \ac{BMC} via \ac{IPMI}. This breaks the attack chain though firmware upgrade. Secondly, \ac{I2C} packet filtering has been implemented in the \ac{BMC}, which prevents \ac{IPMI} commands to directly send packets to the \ac{PMBus}. Moreover, our \pmdetect tool shows that the \ac{VR} is not connected to the \ac{CPU}, which prevents an attack directly from the operating system. 
In conclusion, to the best of our knowledge, we believe that \tSuperMoboNew is not directly vulnerable to the attacks described in this paper. However, we note that as-of-yet unknown vulnerabilities might remain in the firmware update process and the complex software stack running on the \ac{BMC}, which warrants further investigation. 

\section{Conclusions and Countermeasures}
\label{sec:conclusion}

In this paper, we demonstrated two remote attacks that use the \ac{PMBus} interface to control the CPU voltage. An undervolting attack can be used to inject fault to the \ac{SGX} enclave of the CPU and \eg recover a secret key used in cryptography algorithms. The overvolting attack causes permanent damage to the CPU. 

\changed{The attack affects, to our knowledge, all 11th generation Supermicro systems. It also impacts ASRock (tested with \tAsRockServer), though as described the \ac{VRM} behaves differently to Supermicro. We suspect that the attack might also affect other vendors (given that \acp{BMC} are often similar), but could not further investigate this and thus leave it for future work. }

\subsection{Server Platform Security and Embedded System Security}
We first discuss the security considerations for server platforms. Previous security research on computer platforms were mainly focused on the security of the software (either running on the CPU or the management controller). However, each subsystem on a server platform does not act in isolation. Instead, they may interact with each other via the physical connections on the motherboard. In our attacks, we show that the hardware design of the system with a correctly implemented \texttt{ipmitool} can lead to severe security issues and damage to the system. 

Apart from the components on the motherboard, one should also take ``plugin'' devices into consideration when analysing the security of server platforms. During our investigation of the system, we found that when a \ac{PCI-E} device is plugged onto the motherboard, it is also connected to the \ac{I2C} bus of the motherboard. However, if the firmware of a \ac{PCI-E} device is compromised, it can gain access to the \ac{PMBus} to perform the same attacks described in this paper. On \tSuper, this connection can be configured with a jumper named \texttt{JI2C}. Although this jumper is disconnected by default, the user may not be aware of the security implications of connecting this jumper.

In summary, the server platform is a system that has multiple components and microcontrollers. The security of the platforms is not only down to ensuring the security of the software running on it, but the overall design of the hardware and embedded systems on the motherboard should also go through a thorough security review. Securing such a system needs collaborative effort of both software developers and hardware engineers.

\subsection{SGX Security}
Our attack on \ac{SGX} enclaves shows that a privileged local attacker can inject a fault to the enclave and recover secret information with the server management interface, effectively reviving Plundervolt-like software undervolting attacks on Supermicro X11 motherboards. We also demonstrate that a malicious service provider (\eg cloud hoster) can use the attack chains described in the paper to break the security guarantee provided by \ac{SGX}. Moreover, the vulnerability currently cannot be detected/mitigated by \ac{SGX} attestation, because the \ac{BMC} and its firmware are not within the scope of \ac{SGX} attestation. 

A supply chain attack is also possible: as the firmware is not securely verified, it is possible for a \changed{third party} to implant malware into  the \ac{BMC} and \changed{later} launch remote attacks on \ac{SGX} and/or damage the CPU. \changed{Such a firmware modification is also conceivable while the device is being shipped to the end user.} Detecting such attack would be hard, as the firmware of the \ac{BMC} is stored in a separate flash chip. The software running on the \ac{BMC} is thus usually out-of-scope of traditional malware detection methods. 

\subsection{Countermeasures}

\paragraph{Overvolting Attack}

According to our experiments, \ac{PMBus}-based overvolting can lead to permanent damage to the CPU and thus permanent \ac{DoS} of the system. 

The fundamental issue that leads to this attack is the lack of a hardcoded voltage limit of the \ac{VRM}. Simply adding signature verification of the \ac{BMC} firmware or using secure boot to break the attack chain might not be sufficient to prevent overvolting, as other, future attacks might also yield \ac{PMBus} access. Besides, configuring software-based \ac{PMBus} read/write limitations of the \ac{VRM} through the \texttt{MFR\_PWD\_USER} command is also insufficient to stop the attack. This is because this features only sets a 16-bit passcode, which is prone to brute force attack. 
We suggest the following mitigations be implemented for this attack to break the attack chain:

\begin{enumerate}[leftmargin=*, topsep=1pt,itemsep=-3pt]
    \item In the short term, the user manual of the relevant system(s) should be updated to describe the usage and suggested configuration of the \texttt{SMBDAT\_VRM} and \texttt{SMBCLK\_VRM} jumpers, if they are present on a specific model.
 
    \item In the long term, an alternative \ac{VRM} with a hardwired voltage safety limit should be used to replace the current \ac{VRM}. 
    
    \item Another mitigation would be implementing an \ac{I2C} filter to detect and block malicious \ac{PMBus} packets. \texttt{MFR\_VR\_CONFIG}, which can be used to set a 10\,mV \ac{VID} table, is one of the main commands that need to be blocked. Optionally, other commands that involved in the overclocking procedure could be blocked, however, this may affect users who actually \changed{want} to use this feature. Such a filter could be implemented in a small microcontroller that listens to the \ac{I2C} bus and ``jams'' malicious commands by actively pulling the bus low once the command has been detected but before its transmission has been completed.
\end{enumerate}

\paragraph{\ac{PMBus}-based \ac{SGX} Undervolting}
To the best of our knowledge, \pmattackname represents the first attack that directly breaches integrity guarantees in the Intel \ac{SGX} security architecture through the \ac{PMBus} interface. We believe that the fix currently deployed by Intel against \changed{Plundervolt/V0ltPwn (\pvCVE)}---disabling the \ac{SVID} undervolting interface---is insufficient when a remote attacker can get access to the \ac{PMBus} through the \ac{BMC} or \ac{I2C} interface of the CPU, as is the case for Supermicro X11 motherboards.
We note that there might be many other devices connected to the bus, including \ac{PCI-E} devices like graphic cards. It is thus also possible for a compromised \ac{PCI-E} device to send malicious commands to control the CPU voltage. 

Given the potential impact of our findings regarding  fault injection into \ac{SGX} enclaves, in the short term, we recommend inserting software-based fault injection countermeasures into cryptographic computations in enclaves (\eg the quoting enclave). However, we note that such fixes can only serve as mitigations, but not fully eliminate this attack vector.

We would like to highlight that in our opinion, this attack surface \emph{cannot} be easily addressed by jumpers to disconnect the \ac{VRM} from the \ac{SMBus} or adding signature verification of the \ac{BMC} firmware, as we believe that \ac{SGX} attestation cannot independently verify the relevant system configurations: 
\begin{enumerate}[leftmargin=*, topsep=1pt,itemsep=-3pt]
	\item The existence of a \ac{PMBus}/\ac{SMBus} interface to the \ac{VRM} and whether it can be controlled through the \ac{I2C} interface of the CPU;
	\item The existence of an external microcontroller on the motherboard and if it has the functionality to control the \ac{VRM} (\eg~\ac{BMC} or other \ac{PCI-E} devices);
	\item The firmware security status of the \ac{BMC} and other devices on the \ac{PMBus}.
\end{enumerate}
This will make it impossible to give \ac{SGX} assurance of the trust status of the system.
We believe that in the long term, appropriate hardware countermeasures \emph{inside} the CPU package is required: this could on the one hand include continuous monitoring of the received supply voltage, as recently presented by Intel for critical parts of their systems~\cite{intel-fault}, and on the other the use of fully-integrated voltage regulators.

\section*{Acknowledgements}

This research is partially funded by the Engineering and Physical Sciences Research Council (EPSRC) under grants EP/R012598/1, EP/R008000/1, and EP/V000454/1.  The results feed into DsbDtech. We would also like to thank Supermicro for providing a X12DPi-NT6 server for further investigation of the issue.

\appendix
\section{\texttt{i2cdetect} Result for \tSuperMobo}\label{appdx:i2cdetect}
\begin{lstlisting}
~$ sudo i2cdetect 0
         0  1  2  3  4  5  6  7  8  9  a  b  c  d  e  f
[00-20]: -- -- -- -- -- -- -- -- -- -- -- -- -- -- -- --
30:     -- -- -- -- -- -- -- 37 -- -- -- -- -- -- -- --
40:     -- -- -- -- -- -- -- -- -- -- -- -- -- -- -- --
50:     50 -- -- -- -- -- -- -- 58 -- -- -- -- -- -- --
60:     -- -- -- -- -- -- -- -- -- -- -- -- -- -- -- --
70:     -- -- -- -- -- -- -- --

~$ sudo i2cdetect 1
     0  1  2  3  4  5  6  7  8  9  a  b  c  d  e  f
00:          -- -- -- -- -- 08 -- -- -- -- -- -- --
10: 10 -- -- -- -- -- -- -- -- 19 -- -- -- -- -- --
20: 20 -- -- -- -- -- -- -- -- -- -- -- -- -- -- --
30: 30 -- -- -- -- 35 36 -- -- -- -- -- -- -- -- --
40: -- -- -- -- 44 -- -- -- -- -- -- -- -- -- -- --
50: -- 51 -- -- -- -- -- -- -- -- -- -- -- -- -- --
60: -- -- -- -- -- -- -- -- -- -- -- -- -- -- -- --
70: -- -- -- -- -- -- -- --
\end{lstlisting}

\section{\pmdetect Result for \tSuperMobo}\label{appdx:detect_res}
\begin{lstlisting}
$ sudo modprobe i2c_i801
$ sudo ./pmbusdetect -d /dev/i2c-1
Device 0x20              READ_TEMPERATURE success: 0019
!!!!!!!!!!! Detected! Device addr: 20 !!!!!!!!!!!
Device 0x20              SVID_VENDOR_PRODUCT_ID success, data: 2555
This device is likely to be a MPS VRM
Device 0x20 : 00         READ_PAGE success  # Save the page

Page: 00
Device 0x20 : 00         WRITE_PAGE success
Device 0x20 : 00         READ_VOUT success: 00D8

Page: 01
Device 0x20 : 01         WRITE_PAGE success
Device 0x20 : 01         READ_VOUT success: 0001
Device 0x20 : 00         WRITE_PAGE success # Restore the page
\end{lstlisting}

\bibliographystyle{alpha}
\bibliography{abbrev3,crypto,biblio}

\end{document}